\documentclass[lettersize,journal]{IEEEtran}

\def\BibTeX{{\rm B\kern-.05em{\sc i\kern-.025em b}\kern-.08em
    T\kern-.1667em\lower.7ex\hbox{E}\kern-.125emX}}

\usepackage[backend=biber, doi=false,isbn=false,eprint=false,maxbibnames=99,sorting=none]{biblatex}

\usepackage[breaklinks,colorlinks,linkcolor=blue]{hyperref}
\usepackage[usenames,dvipsnames]{xcolor}
\hypersetup{citecolor=blue,linkcolor=blue}
\usepackage{epsfig}
\usepackage{balance} 
\usepackage{color}
\usepackage{comment}
\usepackage{epic}
\usepackage{epsf}
\usepackage{listings}
\usepackage{makecell}
\usepackage{multirow}
\usepackage{textcomp}
\usepackage{xspace}    
\usepackage{latexsym}
\usepackage{afterpage}
\usepackage{amstext}   
\usepackage{amsmath}
\usepackage{graphicx}
\usepackage{tabularx}
\usepackage[caption=false]{subfig}
\usepackage{longtable}
\usepackage{rotating}
\usepackage[algoruled, linesnumbered]{algorithm2e}
\usepackage{fp}
\usepackage{siunitx}
\usepackage{xstring}
 
\usepackage{multirow}
\usepackage{tabularx}
\usepackage{ragged2e}

\usepackage{booktabs}

\usepackage{calculator}
\usepackage{calculus}
\usepackage{siunitx} 
\usepackage{numprint}
\usepackage{enumitem}
\usepackage{threeparttable}
\usepackage{pifont}

\usepackage{algpseudocode}

\DeclareFieldFormat{url}{[Online]. Available\addcolon\space\url{#1}}

\newcommand{\Norothead}[2][0]{\makebox[9mm][c]{\rotatebox{#1}{\makecell[c]{#2}}}}
\newlength\tbspace
\newcolumntype{C}{c<{\hspace{\tbspace}}}

\SetCommentSty{mycommfont}

\makeatletter
\newcommand\footnoteref[1]{\protected@xdef\@thefnmark{\ref{#1}}\@footnotemark}
\makeatother
\newcommand{\quotes}[1]{``#1''}
\newcommand{\DefMacro}[2]{\expandafter\newcommand\csname rmk-#1\endcsname{#2}}
\newcommand{\UseMacro}[1]{\csname rmk-#1\endcsname}

\newcommand{\Space}[1]{}

\makeatletter
\newcommand{\labitem}[2]{%
\def\@itemlabel{\textbf{#1}}
\item
\def\@currentlabel{#1}\label{#2}}

\SetKwProg{Fn}{Function}{}{}

\newcommand{\PP}[1]{
\vspace{2px}
\noindent{\bf{#1.}}
}
\renewcommand{\paragraph}[1]{\smallskip\noindent\emph{#1}\quad}

\newcommand{\Vc}{\pmb{\checkmark\kern-1.1ex\raisebox{.7ex}{\rotatebox[origin=c]{125}{--}}}}

\newcommand{\Sys}{HADES\xspace} 

\newcommand{\bigEI}[1]{$\mathcal{E}_{#1}$\xspace}

\newcommand{\bigAI}[1]{$\mathcal{A}_{#1}$\xspace}
\newcommand{\bigRI}[1]{$\mathcal{R}_{#1}$\xspace}

\SetCommentSty{algocommfont}

\algnewcommand{\algorithmicgoto}{\textbf{goto}}%
\algnewcommand{\Goto}{\algorithmicgoto\xspace}%

\usepackage{tikz}

\newcommand\submittedtext{%
  \footnotesize This work has been submitted to the IEEE for possible publication. Copyright may be transferred without notice, after which this version may no longer be accessible.}

\newcommand\submittednotice{%
\begin{tikzpicture}[remember picture,overlay]
\node[anchor=south,yshift=10pt] at (current page.south) {\fbox{\parbox{\dimexpr0.65\textwidth-\fboxsep-\fboxrule\relax}{\submittedtext}}};
\end{tikzpicture}%
}
\DefMacro{APT29_Scenario2_HADES_Stage1_FP}{60}
\DefMacro{APT29_Scenario2_HADES_Stage1_FN}{0}
\DefMacro{APT29_Scenario2_HADES_Stage2_FP}{0}
\DefMacro{APT29_Scenario2_HADES_Stage2_FN}{0}

\DefMacro{APT29_Scenario2_Elastic_FP}{427}
\DefMacro{APT29_Scenario2_Elastic_FN}{1}
\DefMacro{APT29_Scenario2_Elastic_Lat_Mov_FP}{148}
\DefMacro{APT29_Scenario2_Elastic_Lat_Mov_FN}{1}
\DefMacro{APT29_Scenario2_Chronicle_FP}{19}
\DefMacro{APT29_Scenario2_Chronicle_FN}{1}
\DefMacro{APT29_Scenario2_Chronicle_Lat_Mov_FP}{14}
\DefMacro{APT29_Scenario2_Chronicle_Lat_Mov_FN}{1}
\DefMacro{APT29_Scenario2_Sigma_FP}{812}
\DefMacro{APT29_Scenario2_Sigma_FN}{0}
\DefMacro{APT29_Scenario2_Sigma_Lat_Mov_FP}{545}
\DefMacro{APT29_Scenario2_Sigma_Lat_Mov_FN}{0}

\DefMacro{WizardSpider_HADES_Stage1_FP}{66}
\DefMacro{WizardSpider_HADES_Stage1_FN}{0}
\DefMacro{WizardSpider_HADES_Stage2_FP}{2}
\DefMacro{WizardSpider_HADES_Stage2_FN}{0}

\DefMacro{WizardSpider_Elastic_FP}{519}
\DefMacro{WizardSpider_Elastic_FN}{1}
\DefMacro{WizardSpider_Elastic_Lat_Mov_FP}{194}
\DefMacro{WizardSpider_Elastic_Lat_Mov_FN}{1}
\DefMacro{WizardSpider_Chronicle_FP}{45}
\DefMacro{WizardSpider_Chronicle_FN}{1}
\DefMacro{WizardSpider_Chronicle_Lat_Mov_FP}{30}
\DefMacro{WizardSpider_Chronicle_Lat_Mov_FN}{1}
\DefMacro{WizardSpider_Sigma_FP}{1173}
\DefMacro{WizardSpider_Sigma_FN}{0}
\DefMacro{WizardSpider_Sigma_Lat_Mov_FP}{605}
\DefMacro{WizardSpider_Sigma_Lat_Mov_FN}{0}

\DefMacro{Oilrig_HADES_Stage1_FP}{156}
\DefMacro{Oilrig_HADES_Stage1_FN}{0}
\DefMacro{Oilrig_HADES_Stage2_FP}{2}
\DefMacro{Oilrig_HADES_Stage2_FN}{0}

\DefMacro{Oilrig_Elastic_FP}{450}
\DefMacro{Oilrig_Elastic_FN}{0}
\DefMacro{Oilrig_Elastic_Lat_Mov_FP}{213}
\DefMacro{Oilrig_Elastic_Lat_Mov_FN}{0}
\DefMacro{Oilrig_Chronicle_FP}{237}
\DefMacro{Oilrig_Chronicle_FN}{1}
\DefMacro{Oilrig_Chronicle_Lat_Mov_FP}{37}
\DefMacro{Oilrig_Chronicle_Lat_Mov_FN}{1}
\DefMacro{Oilrig_Sigma_FP}{791}
\DefMacro{Oilrig_Sigma_FN}{0}
\DefMacro{Oilrig_Sigma_Lat_Mov_FP}{417}
\DefMacro{Oilrig_Sigma_Lat_Mov_FN}{0}

\DefMacro{APT29_Scenario2_Memory_Usage_min}{0}
\DefMacro{WizardSpider_Memory_Usage_min}{0}
\DefMacro{Oilrig_Memory_Usage_min}{0}

\DefMacro{APT29_Scenario2_Memory_Usage_max}{160}
\DefMacro{WizardSpider_Memory_Usage_max}{125}
\DefMacro{Oilrig_Memory_Usage_max}{249}

\DefMacro{APT29_Scenario2_Memory_Usage_mean}{141}
\DefMacro{Sandworm_Scenario1_Memory_Usage_mean}{}
\DefMacro{WizardSpider_Memory_Usage_mean}{116}
\DefMacro{Oilrig_Memory_Usage_mean}{165}

\DefMacro{fpr_avg_reduction}{98}
\DefMacro{fpr_avg_1to2_reduction}{99}

\begin{document}

\title{HADES: Detecting Active Directory Attacks via Whole Network Provenance Analytics}

\author{Qi Liu, Kaibin Bao, Wajih Ul Hassan, Veit Hagenmeyer
\thanks{This work was supported by funding of the Helmholtz Association (HGF) through the Energy System Design (ESD) program. 

Qi Liu, Kaibin Bao, and Veit Hagenmeyer are with Institute for Automation and Applied Informatics, Karlsruhe Institute of Technology (KIT), Eggenstein-Leopoldshafen 76344, Germany (e-mail: qi.liu@kit.edu; kaibin.bao@kit.edu; veit.hagenmeyer@kit.edu).

Wajih Ul Hassan is with the School of Engineering \& Applied Science, University of Virginia, Charlottesville, VA 22904-4740, USA (e-mail: hassan@virginia.edu).
}}

\maketitle

\submittednotice 

\begin{abstract}
Due to its crucial role in identity and access management in modern enterprise networks, Active Directory (AD) is a top target of Advanced Persistence Threat (APT) actors. Conventional intrusion detection systems (IDS) excel at identifying malicious behaviors caused by malware, but often fail to detect stealthy attacks launched by APT actors. Recent advance in provenance-based IDS (PIDS) shows promises by exposing malicious system activities in causal attack graphs. However, existing approaches are restricted to intra-machine tracing, and unable to reveal the scope of attackers’ traversal inside a network. We propose \Sys, the first PIDS capable of performing accurate causality-based cross-machine tracing by leveraging a novel concept called \textit{logon session based execution partitioning} to overcome several challenges in cross-machine tracing. We design \Sys as an efficient on-demand tracing system, which performs whole-network tracing only when it first identifies an authentication anomaly signifying an ongoing AD attack, for which we introduce a novel lightweight authentication anomaly detection model rooted in our extensive analysis of AD attacks. To triage attack alerts, we present a new algorithm integrating two key insights we identified in AD attacks. Our evaluations show that \Sys outperforms both popular open-source detection systems and a prominent commercial AD attack detector.

\end{abstract}


\section{Introduction}
Recent CrowdStrike studies~\cite{CrowdStrikeStudy2023,CSTH2023,Shastri2022,CSblog2023} show that 80\% of modern cyberattacks are identity-driven, i.e., leveraging compromised credentials. After the initial access, attackers increasingly target Microsoft Active Directory for obtaining critical domain credentials, and hence gaining the ability to move laterally in the victim network and escalate privilege. For instance, Kerberoasting attacks~\cite{T15583} increased almost 600\% from 2022 to 2023, and Pass-the-Hash attacks~\cite{T15502} increased 200\%~\cite{CSTH2023}.
   
Active Directory (Domain Service) plays a crucial role for identity and access management in modern enterprises' IT infrastructures, with 90\% of Fortune 1000 companies relying on it~\cite{FrostSullivanSurvey,CSblog2023}. To advance the attack after the initial access, attackers routinely leverage Active Directory (AD) functionalities in several stages of the cyber-kill-chain, in particular, internal reconnaissance, credential access, lateral movement and privilege escalation. Comparing to network scanning using third-party tools like \texttt{nmap}~\cite{nmap}, SPN (Service Principle Name) scanning, a form of AD internal reconnaissance, via native Windows programs like \texttt{setspn}~\cite{setspn} is much less noisy while achieving the same goal, i.e., identifying potentially vulnerable servers in the network. Hence, advanced attackers like APT (Advanced Persistent Threat) actors primarily rely on Living-Off-the-Land Binaries (LOLBins)~\cite{LOLBins2, TrellixStudy2023}, complicating intrusion detection and posing significant risks to enterprises.

Conventional intrusion detection systems (IDS) often focus on isolated system events, and excel at identifying malware causing a burst of malicious behaviors. Facing APT actors frequently employing LOLBins and the so-called low and slow strategy, these IDS tend to miss those attacks and suffer from a high false negative rate. To overcome this, security vendors resort to the MITRE ATT\&CK Matrix~\cite{mitrematrix} as a reference for further creating detection rules. The MITRE ATT\&CK Matrix, maintained with contributions from leading industrial security vendors, provides a high-level summary of attack tactics and techniques, including usage of LOLBins. Our evaluation in Section~\ref{s:eval} shows that the isolated analysis of these IDS create an excessive amount of false alerts due to the fact that LOLBins are programs executed frequently also by normal users.

Provenance-based intrusion detection systems (PIDS)~\cite{protracer,hossain2017sleuth, winnower2018,holmes2019,nodoze2019,rapsheet2020,hossain2020combating,KAIROS,FLASH,shadewatcher,PROGRAPHER, dong2023industrial} emerged as a new solution, stitching causally related system activities by parsing system events like process creation, file access and network socket access into provenance graphs. Given a suspicious event as a starting point, a backward tracing and forward tracing in the provenance graph can expose more related malicious system activities caused by attackers, i.e., the root cause and attack ramifications, respectively. PIDS have proved to be effective in particular in reducing false alarm rates and presenting real attack activities in attack graphs. Attack graphs provide much richer context for security analysts to further investigate the scope of an attack, invaluable for an effective and efficient security operation center (SOC).  

\begin{figure*}[!t]
  \centering
  \includegraphics[width=0.7\linewidth]{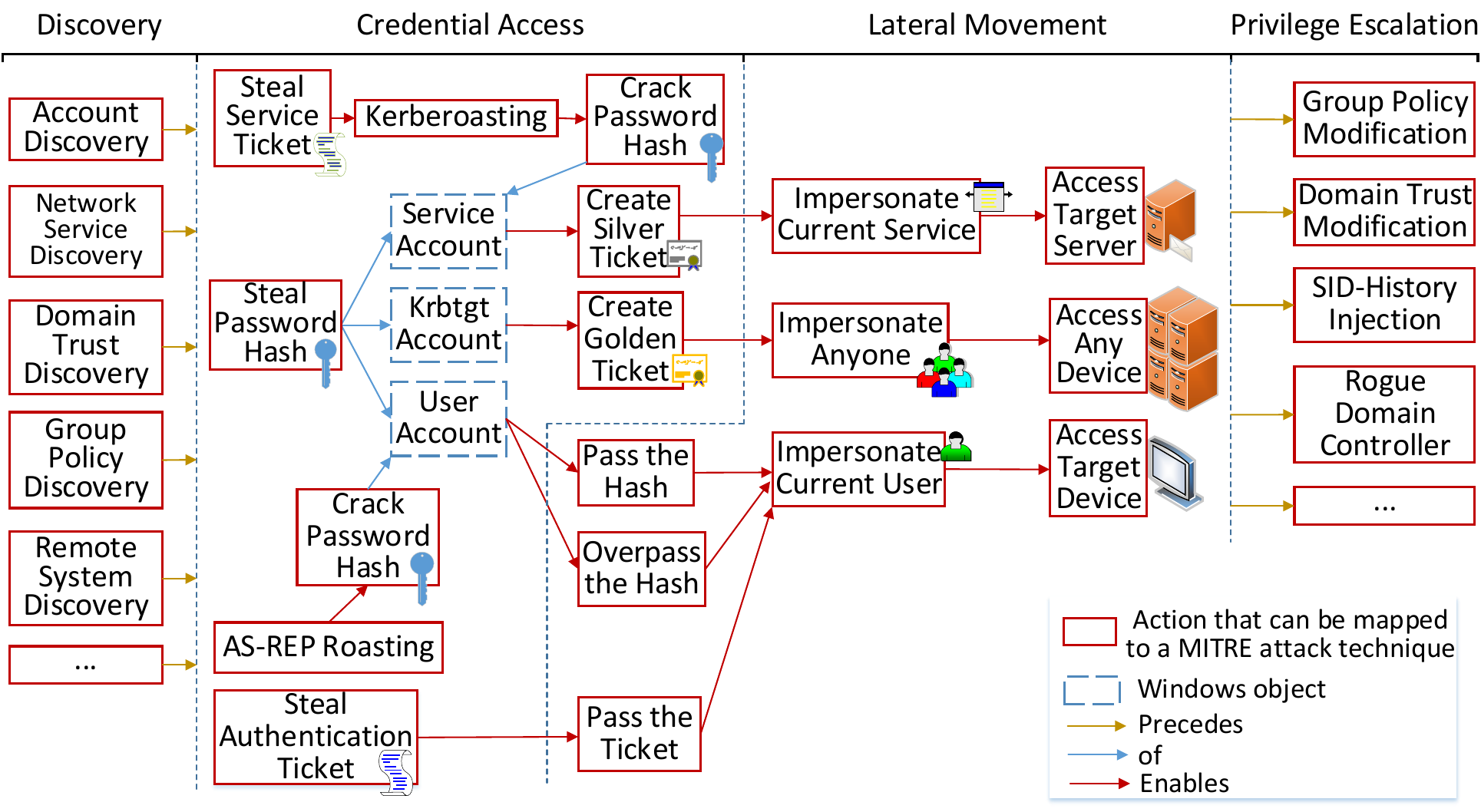}
  \caption{Active directory attack overview.}
    \vspace{-1ex}
    \label{fig:ADMapping}
\end{figure*}

However, current PIDS are constrained to intra-machine provenance tracing, and lack the ability to track across machines in an enterprise environment, and accurately reveal the scale of attackers' traversal inside the network, crucial for attack remediation. Cross-machine provenance tracing faces significant challenges due to the notorious dependency explosion problem~\cite{beep,ma2015accurate}. While intra-machine dependency explosion occurs for long-running processes, in which each input is conservatively considered causally responsible for all subsequent outputs, and vice versa, cross-machine dependency explosion arises if cross-machine edges are created simply on a network connection basis. Naively connecting two intra-machine provenance graphs, whenever there is a logon event from one machine to another, or even whenever there is a network connection between them~\cite{Trace2021}, would inevitably result in numerous false dependencies, as discussed further in Section~\ref{ss:logonSessionBasedTracing} in detail. Recognizing the critical importance of logon session ID, we propose a novel concept named \textit{logon session based execution partitioning and tracing}. By leveraging both authentication \& logon logs and system logs, our system \Sys is capable of 1) fine-grained cross-machine provenance tracing, 2) alleviating dependency explosion also for intra-machine provenance tracing, 3) drastically reducing log size, 4) automatically pinpointing privilege escalation.

\Sys employs a two-stage approach for efficient AD attack detection. Its first stage consists of a light-weight authentication anomaly detection model responsible for identifying potential AD attacks and forwarding its results to \Sys's stage two component, which performs logon session-based tracing and attack graph triage. Both our authentication anomaly detection model and attack graph triage algorithm are rooted in our thorough analysis of AD attacks. 
Through \Sys's development, we faced several difficulties in realizing accurate cross-machine tracing. First, depending on remote access type, each authentication \& logon process causes varying number of logon events with distinct logon session ID, complicating the identification of the correct session ID. Second, under certain circumstances, system activities in a new logon session are assigned with an existing session ID, causing false dependencies. Third, it is often not possible to disclose the remote access type by examining the current logon event alone. We overcome these challenges by introducing a remote access type inference module, a logon session ID reassignment module, and a logon session linking module in \Sys, based on our extensive profiling and analysis of Windows logging frameworks.  Our implementation of logon session-based tracing avoids time-consuming instrumentation, error-prone training, and applies to the whole system rather than a single program.

Unlike previous techniques, \Sys produces alarms satisfying all five properties of reliability, explainability, analytical depth, contextuality, and transferability as introduced in \cite{Alahmadi2022}. In its first stage, \Sys's anomaly detection model avoids easily changeable indicators such as hard-coded IP addresses and file hashes, which are typical in popular Security Information and Event Management (SIEM) detection rules~\cite{elasticdetectionrules, sigma}, ensuring reliable detection. The attack graphs produced in the second stage of \Sys are both explainable and contextual, providing an analytical overview of the attack. Finally, the system's customizable weighting factors for indicators introduced in its threat score calculation make \Sys highly transferable and adaptable for practical use in various scenarios.

The main contributions of this paper are as follows:
\begin{itemize}[noitemsep,topsep=0pt]
 \item We give a succinct and contextualized AD attack overview based on a thorough analysis, critical for understanding and detecting AD attacks.
 \item We present a light-weight authentication anomaly detection model for AD attacks.
 \item We propose a novel concept called logon session-based execution partitioning and tracing.
 \item We demonstrate the first accurate and efficient causality-based cross-machine provenance tracing system \Sys.
 \item We introduce a new alert triage algorithm for accurate AD attack detection.
\end{itemize}

\section{Background}
\label{s:background}

AD attacks are a set of cyberattacks targeting Microsoft Active Directory service employed in most corporate environments. Half of organizations worldwide have experienced an AD attack in recent years, while 40\% of those attacks were successful~\cite{securityboulevard,CSblog2023}. AD uses a database to store critical information about organizations' network resources, and provides administrators the ability to manage the access to those resources. That is, AD holds the blueprint of an organization's environment, and the keys to all available resources. Consequently, various kinds of attacks specifically exploiting vulnerabilities in Microsoft's AD design and implementation emerged over time, resulting in a partial or even full-scale compromise of many enterprise networks~\cite{Talyanski2022,Tenable2021}.

Often, an attack is launched first against a targeted user via spear-phishing, or a vulnerable domain-joined machine accessible from outside, to get an initial foothold inside the domain. The attacker then performs internal AD discovery to gain knowledge about the internal network and spot more machines, before moving to them typically via some form of stolen credentials from the first machine. 

A proper understanding of the prerequisites of various AD attacks, and logging possibilities as well as limitations is critical for accurate AD attack detection. Figure~\ref{fig:ADMapping} shows the AD attack skeleton, in which most relevant AD attack techniques are listed. This model serves as a blueprint for reliably detecting AD attack techniques and reconstructing high-level AD attack graphs. These graphs can be further expanded to create whole network attack graphs, including all malicious system activities conducted by attackers.

\section{Threat Model}

Like other PIDS~\cite{unicorn2020,protracer,hossain2017sleuth, winnower2018,holmes2019,nodoze2019,rapsheet2020,hossain2020combating,KAIROS,FLASH,shadewatcher,PROGRAPHER}, our system \Sys assumes the integrity of the underlying operating systems, firmware and our logging frameworks. We only consider attacks in which threat actors already achieved an initial foothold on a domain-joined host via binary exploitation or social engineering, and move laterally to other machines leveraging stolen credentials instead of program exploitation. Attacks restricted to only one machine are deemed less severe and are hence out of \Sys's detection scope.

\section{System Design}
\label{s:design}

\begin{figure}[!t]
  \centering
  \includegraphics[width=\linewidth]{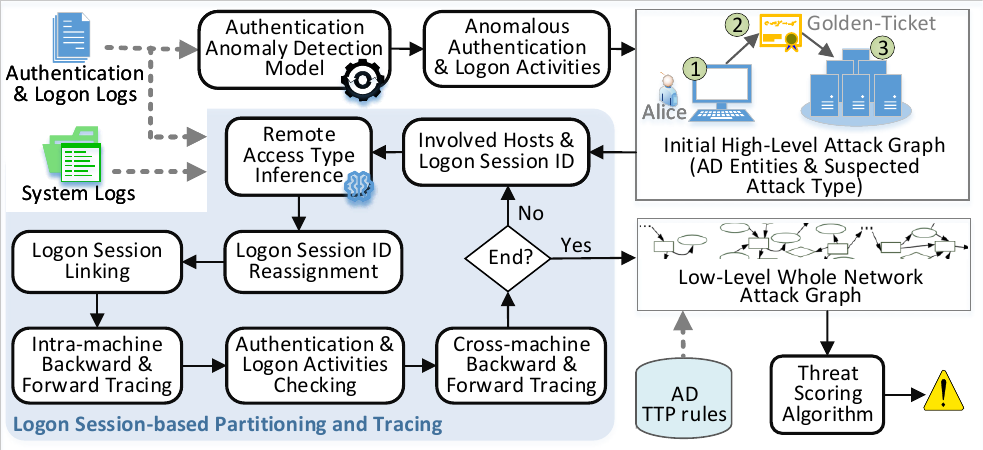}
  \caption{\Sys overview.}
    \vspace{-1ex}
    \label{fig:SystemOverview}
\end{figure}

\begin{algorithm}[!tbp]
  \scriptsize
  \DontPrintSemicolon
  \SetKwInOut{Input}{Inputs}
  \SetKwInOut{Output}{Output}
  
  \Input{System audit log events \bigEI{}; \\
    Authentication \& logon events \bigAI{};\\
    AD TTP rules \bigRI{}}
  \Output{List $L_{<AG,TS>}$ of attack graph and its threat score pairs} 
  \BlankLine
  
  \SetKwFunction{GetAG}{\sc{GetADAttackGraph}}
  \SetKwFunction{GetAA}{\sc{GetAuthenticationAnomaly}}
  \SetKwFunction{CreateHG}{\sc{CreateHighLevelGraph}}
  \SetKwFunction{GetRAT}{\sc{CheckRemoteAccessType}}
  \SetKwFunction{ReassignSID}{\sc{ReassignSessionID}}
  \SetKwFunction{LinkSes}{\sc{LinkSessions}}
  \SetKwFunction{GetSc}{\sc{GetThreatScore}}
    
  \Fn{\GetAG{\bigEI{}, \bigAI{}}}{%
  \tcc{Get a list of authentication \& logon anomalies}
   $L_{{<ae_{\gamma}, \mathcal{L{\gamma}}, \mathcal{T{\gamma}}>}}$ $\leftarrow$ \GetAA(\bigAI{})  \label{line:CheckAuthAno}
    
  \ForEach{($\mathcal{A{\gamma}}$, $\mathcal{L{\gamma}}$, $\mathcal{T{\gamma}}$) $\in$ $L_{{<\mathcal{A{\gamma}}, \mathcal{L{\gamma}}, \mathcal{T{\gamma}}>}}$}{     
    $AG$ $\leftarrow$ null 
     
    $HG$ $\leftarrow$ \CreateHG($\mathcal{A{\gamma}}$, $\mathcal{L{\gamma}}$) \label{line:CreateHG}
    
    $AG$ $\leftarrow$ $AG$ $\cup$ $HG$
    
    $AccessType$ $\leftarrow$ \GetRAT($\mathcal{A{\gamma}}$, \bigEI{})  \label{line:GetRAT}
    
    $SessionID$ $\leftarrow$ \ReassignSID($AccessType$, $\mathcal{A{\gamma}}$, \bigEI{}) 
    
    $LinkedSessionID$ $\leftarrow$ \LinkSes($SessionID$, $\mathcal{A{\gamma}}$, \bigEI{})
    
    ($\mathcal{E{\alpha}}$, $G{\alpha}$) $\leftarrow$ \textsc{TraverseBackward}($SessionID$, $LinkedSessionID$, \bigEI{}) \label{line:TraversalBackwardSystem}
    
    ($\mathcal{E{\kappa}}$, $G{\kappa}$) $\leftarrow$ \textsc{TraverseForward}($SessionID$, $LinkedSessionID$, \bigEI{})\label{line:TraversalForwardSystem}
    
    $AG$ $\leftarrow$ $AG$ $\cup$ $G{\alpha}$ $\cup$ $G{\kappa}$
    
    $\mathcal{E{\alpha}}$ $\leftarrow$ $\mathcal{E{\alpha}}$ $\cup$ $\mathcal{E{\kappa}}$
    
    $\mathcal{A{\alpha}}$ $\leftarrow$ \textsc{CheckAuthenticationLogon}($\mathcal{E{\alpha}}$, \bigAI{})
    
    \If{$\mathcal{A{\alpha}}$ is not null} {
           \Goto \ref{line:GetRAT}
     }
     
     $TS$ $\leftarrow$ \GetSc($AG$)  \label{line:GetSc}
     
     $L_{<AG, TS>}$ $\leftarrow$ $L_{<AG, TS>}$ $\cup$ ($AG$, $TS$) 
   
   }
   \Return $L_{<AG, TS>}$
  }
  
  \BlankLine
  \BlankLine

  \Fn{\GetAA{\bigAI{}}}{%
    \ForEach{$ae$ $\in$ \bigAI{}}{
          \bigAI{\alpha} $\leftarrow$ \textsc{TraversalBackward}($ae$) \label{line:TraversalBackwardAuthentication}
          
          \bigAI{\kappa} $\leftarrow$ \textsc{TraversalForward}($ae$) \label{line:TraversalForwardAuthentication}
          
          \bigAI{\alpha} $\leftarrow$ \bigAI{\alpha} $\cup$ \bigAI{\kappa}   
     
          ($\mathcal{L{\alpha}}$, $\mathcal{T{\alpha}}$) $\leftarrow$ \textsc{CheckAttackType}(\bigAI{\alpha}) \label{line:CheckAttackType}
             
         \If{$\mathcal{L{\alpha}}$ $\in$ $L_{{<\mathcal{L}_{attack}>}}$} {
          $L_{{<\mathcal{A{\gamma}}, \mathcal{L{\gamma}}, \mathcal{T{\gamma}}>}}$ $\leftarrow$  $L_{{\mathcal{A{\gamma}}, \mathcal{L{\gamma}}, \mathcal{T{\gamma}}>}}$ $\cup$ ($\mathcal{A{\alpha}}$, $\mathcal{L{\alpha}}$, $\mathcal{T{\alpha}}$)
     } 
   }
   \Return $L_{{<\mathcal{A{\gamma}}, \mathcal{L{\gamma}}, \mathcal{T{\gamma}}>}}$
  }
  
  \BlankLine
  \BlankLine

  \Fn{\GetSc{$AG$}}{%
  
      \ForEach{($CrossMachineEdge$, $\mathcal{T{\omega}}$) $\in$ $AG$}{    
      \tcc{Get a list of discovery techniques and frequency} 
      $L_{{<tec_d, freq>}}$ $\leftarrow$ \textsc{CheckADDiscovery}($AG$, $\mathcal{T{\omega}}$, \bigRI{})
          
      $L_{{<tec_{ca}, freq>}}$ $\leftarrow$ \textsc{CheckCredentialAccess}($AG$, $\mathcal{T{\omega}}$, \bigRI{})
      
      $L_{{<pe>}}$ $\leftarrow$ \textsc{CheckPrivilegeEscalation}($AG$)
          
      $TS_{sub}$ $\leftarrow$ \textsc{CalculateScore}($L_{{<tec_d, freq>}}$, $L_{{<tec_{ca}, freq>}}$, $L_{{<pe>}}$) 
       
       $L_{{<TS_{sub}>}}$ $\leftarrow$  $L_{{<TS_{sub}>}}$  $\cup$  $TS_{sub}$ 
   }  
   
   $TS$ $\leftarrow$ \textsc{SumScore}($L_{{<TS_{sub}>}}$)
   
   \Return $TS$
  }

  \caption{\sc{Active Directory Attack Detection}}
  \label{alg:logon_ses_tra}
\end{algorithm}

\begin{figure*}[!t]
  \centering
  \includegraphics[width=0.98\linewidth]{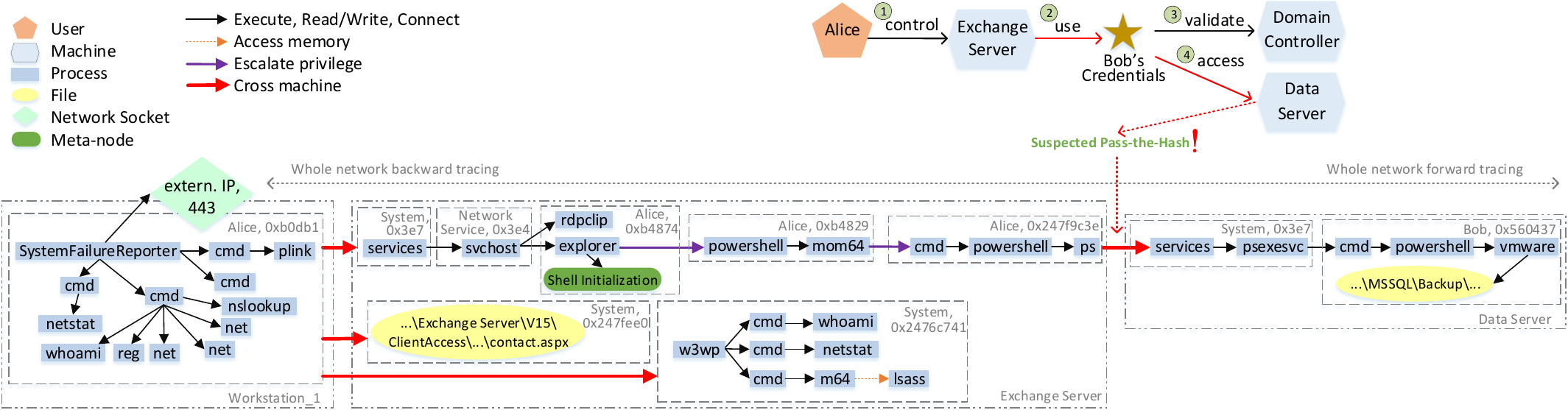}
  \caption{
  An AD attack graph created by \Sys on the Oilrig~\cite{oilrigEP} dataset. \Sys first creates an initial high-level attack graph involving AD entities like users and hosts, after it detects an authentication \& logon anomaly and suspects a Pass-the-Hash attack. Then it performs system-level forward tracing inside the specific logon session under user Bob in the accessed host Data Server, and system-level forward \& backward tracing inside the logon session of Alice in the accessing host Exchange Server. Subsequently, it traces back to a logon session of Alice in the Workstation\_1. Next, it traces forward \& backward inside this logon session, leading to another two logon sessions in the Exchange Server. This graph reveals that an attacker leveraged a C2 (Command \& Control) agent disguised under the process name SystemFailureReporter on the Workstation\_1 to perform AD discovery via LOLBins like \texttt{net} and \texttt{netstat}. The attacker then pivoted to the Exchange Server, performed further AD discovery, and conducted credential access, before moving to the Data Server for accessing critical data.\protect\footnotemark}
    \vspace{-2ex}
    \label{fig:OilrigAttackGraph}
\end{figure*}

Our system \Sys employs a two-stage approach for efficient and accurate detection of AD attacks, producing an initial high-level attack graph and a low-level whole network attack graph, respectively. Figure~\ref{fig:SystemOverview} provides an overview of \Sys, whereas a formal description of \Sys's detection procedure is given in Algorithm~\ref{alg:logon_ses_tra}. In its first stage, \Sys relies on authentication \& logon logs only, and parses through each AD authentication \& logon log event. It traces backward on each logon event, and traces backward \& forward on each authentication event (Algorithm~\ref{alg:logon_ses_tra} Lines~\ref{line:TraversalBackwardAuthentication}-\ref{line:TraversalForwardAuthentication}), as authentication is a multiple-step process. The core component of \Sys's stage 1 is a light-weight authentication anomaly detection model, against which the authentication \& logon tracing result is matched. Once an anomaly is found, \Sys produces a high-level attack graph including involved AD entities (Algorithm~\ref{alg:logon_ses_tra} Line~\ref{line:CreateHG}), i.e., users and machines, and labeled with a suspected AD attack type (Algorithm~\ref{alg:logon_ses_tra} Line~\ref{line:CheckAttackType}), at the end of its stage 1. A concrete example of such high-level attack graphs is given in Figure~\ref{fig:OilrigAttackGraph}, in which it shows that user Alice from the Exchange Server has used user Bob's password hash to authenticate against the Domain Controller, and then accessed the Data Server, mimicking the Pass-the-Hash behavior.

The identified host names, user names and their logon session ID are passed to \Sys's stage 2, in which it performs logon session-based execution partitioning and tracing. On the accessed host, e.g., the Data Server in Figure~\ref{fig:OilrigAttackGraph}, by leveraging both authentication \& logon logs and system logs, \Sys first infers the remote access type, which is critical for deciding whether the logon session ID needs to be reassigned in the next step. Then it links related logon sessions resulted from the same identity, and performs intra-machine backward \& forward tracing inside these logon sessions (Algorithm~\ref{alg:logon_ses_tra} Lines~\ref{line:TraversalBackwardSystem}-\ref{line:TraversalForwardSystem}). Afterwards, \Sys checks the authentication \& logon logs to spot any logon event inside any domain-joined machine initiated from the accessed host, i.e., cross-machine forward tracing. Meanwhile, \Sys examines the authentication \& logon logs to find out whether and from which domain-joined machine the current logon session inside the accessing host, e.g., the Exchange Server in Figure~\ref{fig:OilrigAttackGraph}, is initiated, i.e., cross-machine backward tracing. After the logon session-based tracing ends as no further involved machine can be found, \Sys passes the low-level whole network provenance graph to its threat scoring algorithm (Algorithm~\ref{alg:logon_ses_tra} Line~\ref{line:GetSc}), the final step of stage 2. In this step, we leverage open-source TTP (Tactics, Techniques, Procedures) detection rules~\cite{sigma} for AD discovery and credential access.

\footnotetext{Note that the bounding boxes with a session ID label or host name are manually added for better readability, this information is encoded in nodes in the original attack graph. Besides, we replaced the user names in the original emulation plan with shorter names. Some processes, e.g., ps and vmware, in the attack graph are malicious processes disguised under a false name. Further, we introduced a graph optimization technique called meta-nodes to conclude system activities of standard known benign routines.} 

In the example of Figure~\ref{fig:OilrigAttackGraph}, the whole network forward tracing from the accessed host, i.e., the Data Server, did not lead to further logon session on another machine, as the attacker found the targeted data and did not advance further in the network. However, the whole network backward tracing from the accessing host, i.e., the Exchange Server, reveals that the current logon session on it is a RDP (Remote Desktop Protocol) session that was initiated from another logon session of user Alice inside another machine Workstation\_1. Backward tracing from the Workstation\_1 did not lead to further domain-joined machine, but rather an external IP address, which belongs to the C2 (Command \& Control) server operated by the attacker. Forward tracing from the Workstation\_1 resulted in another two logon sessions in the Exchange Server preceding the RDP logon session on it. In one of these logon sessions, credential access was performed, contributing to the later lateral movement from the RDP logon session on the Exchange Server to a logon session on the Data Server.  Note that backward tracing can only lead to exactly one logon session, while forwarding tracing can lead to multiple logon sessions, as one may remotely access multiple machines from one machine and multiple logon sessions inside one machine do not interfere with each other.

\subsection{Authentication Anomaly Detection}

\begin{figure}[!t]
  \centering
  \includegraphics[width=0.80\linewidth]{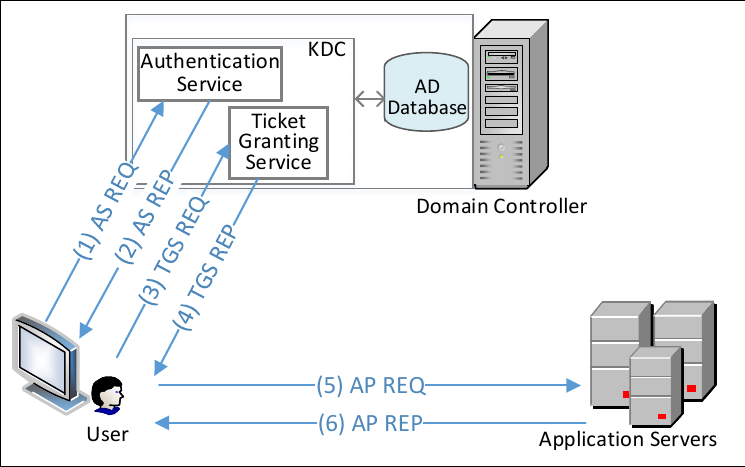}
  \caption{Standard AD authentication process.}
    \vspace{-1ex}
    \label{fig:StandardAuthenticationProcess}
\end{figure}

The main purpose of AD is providing the service of authentication of user or machine accounts and authorization to different network resources. 
AD employs Kerberos as its default authentication protocol, and implements it in two components in a domain controller (DC), i.e., the Authentication Service (AS) and the Ticket-Granting Service (TGS). Figure~\ref{fig:StandardAuthenticationProcess} shows a simplified version of standard AD authentication process: 1) the user sends an encrypted AS request for a TGT (Ticket-Granting-Ticket), 2) the DC replies with a TGT if it can decrypt the request and hence verify the user's identity, 3) the user asks for a TGS ticket by providing the TGT, 4) the DC replies with a TGS ticket, 5) the user asks for access to an application server by providing the TGS ticket, 6) the server gives the user requested access after validating the TGS ticket. The entire process uses shared secret cryptography to deter network-level eavesdropping and replay attacks, meaning that credential access is only possible inside hosts.

The complicated network authentication process is often exploited by attackers who manage to steal some form of credentials or credentials-related data on one machine, i.e., credential access, and ultimately use them to successfully authenticate against a domain controller and hence access another machine, i.e., lateral movement. Several AD attack techniques exhibit similarities, often causing confusion. However, we believe that a practical detection system needs to differentiate between these attacks and hence triage the alerts due to the varying degrees of severity of those attacks. For instance, a Golden-Ticket~\cite{GoldenTicket} attack is more critical than other attack types since it can lead to a full-scale domain compromise.

\begin{figure}[!t]
  \centering
  \includegraphics[width=0.95\linewidth]{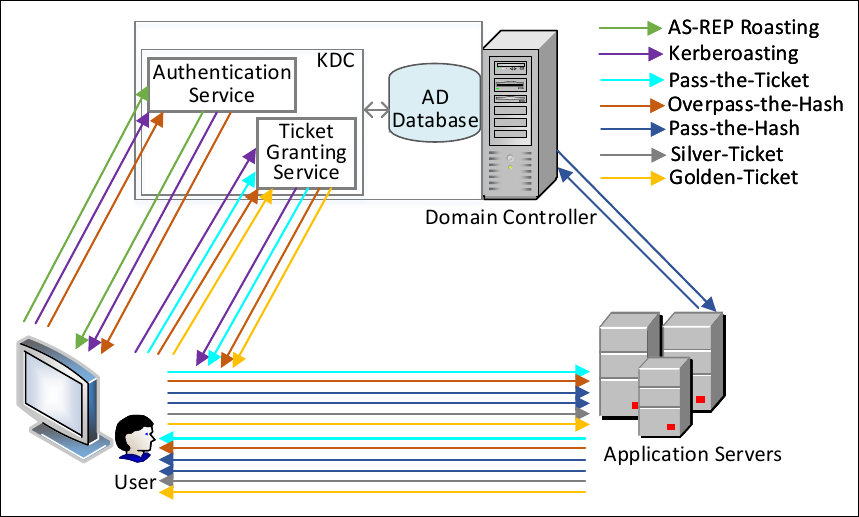}
  \caption{Authentication incompleteness/abnormality.}
    \vspace{-1ex}
    \label{fig:AuthenticationIncompleteness}
\end{figure}

We analyzed each attack type, and identified anomaly in the corresponding authentication process, as illustrated in Figure~\ref{fig:AuthenticationIncompleteness}. For instance, in AS-REP Roasting attacks, attackers only send a TGT request and aim to crack the user's password from the received TGT offline. Hence no TGS request follows. In Kerberosting attacks, attackers aim to crack a service account's password from a received TGS, meaning that no access on the application server follows the TGS request. In Pass-the-Ticket attacks, attackers send a TGS request to the DC with a stolen TGT, meaning that no TGT request precedes the TGS request. 

Therefore, we propose a light-weight authentication anomaly detection model based on our finding. In our evaluation in section~\ref{s:eval}, this model proves to be effective in catching each attack instance, but plagued with a high false positive rate, hindering its adaptability in practice. The reasons for false positives are manifold. For instance, false positives for AS-REP Roasting and Kerberosting can be resulted from network disruption. Further, legitimate use of native Windows programs like \texttt{runas} causes false Pass-the-Hash alerts.

\subsection{Logon Session-based Execution Partitioning and Tracing}
\label{ss:logonSessionBasedTracing}
Recent PIDS have proved to be invaluable in reducing false alarms. To tackle the high rate of false positive, we aim to develop the first causality-based cross-machine PIDS. A naive approach would be to connect the intra-machine provenance graphs of two domain-joined machines with a cross-machine edge, whenever there is a network connection between them, as suggested in~\cite{Trace2021}. However, this kind of cross-machine edges do not represent a causality, but rather a correlation, leading to numerous cross-machine edges between intra-machine provenance graphs. Take Figure~\ref{fig:OilrigAttackGraph} as an example, this would result in 7364 edges between the Exchange Server's provenance graph and other hosts' graphs after only one day operation, while in fact there are only 4 causality-based cross-machine edges to/from the Exchange Server in the true attack chain.

A more advanced approach would be to connect intra-machine provenance graphs, whenever there is a logon event from one machine to another. Yet, due to the prevalence of credential thefts, in particular in AD attacks, it is impossible to identify the true identity behind each logon. For instance, in Figure~\ref{fig:OilrigAttackGraph}, Alice's credentials were stolen by the attacker. Given that both Alice and the attacker have accessed the Exchange Sever from the Workstation\_1 using the same credentials, this approach would still create false-dependencies. In fact, this approach would produce 479 correlation-based cross-machine edges, instead of 4 causality-based, to/from the Exchange Server. Correlation-inferred edges are the root cause of the notorious dependency explosion problem~\cite{beep,ma2015accurate,inam2022sok}. 

In order to enable causality-based provenance tracing, we resort to a critical yet previously undiscovered information: logon session ID, and propose a novel concept: logon session-based execution partitioning and tracing. To the best of our knowledge, this work presents the first PIDS leveraging logon session ID. Windows assigns a logon session ID (unique until next reboot) after each successful authentication \& logon to label system activities run in that session in Windows Security logs~\cite{windowssecurity}. A logon session is a computing session assigned with an access token representing the authenticated account’s security context and permissions~\cite{LsaLogonSessions}. Every Windows process runs within a logon session's context.

\subsubsection{Benefits}
The benefits of logon session-based tracing are fourfold: 1) it enables fine-grained causality-based cross-machine provenance tracing, 2) it alleviates dependency explosion also for intra-machine provenance tracing, 3) it reduces log size as most forensics-irrelevant system activities under logon sessions with predefined logon ID\footnote{Predefined logon ID include \texttt{0x3e4}, \texttt{0x3e5}, and \texttt{0x3e7}, belonging to Network Service account, Local Service account, System account, respectively~\cite{Russinovich2016}.} can be safely discarded, 4) it automatically pinpoints privilege escalation during intra-machine provenance tracing.

When a user from one machine logs into another machine\footnote{Note that a user accessing resources on a remote machine also triggers a logon on the remote machine. The authentication \& logon process is transparent to users due to Single Sign-on.}, the corresponding logon session in the second machine is resulted from exactly one logon session in the first machine. For example, in Figure~\ref{fig:OilrigAttackGraph}, Alice's logon session \texttt{0xb4874} on the Exchange Server is only resulted from Alice's logon session \texttt{0xb0db1} on the Workstation\_1, while there are multiple Alice's logon sessions on the Workstation\_1 in parallel. Our system \Sys can accurately disclose this, as it checks authentication \& logon logs among involved machines and performs logon session-based tracing. \textit{With logon session-based tracing, we can create a whole-network provenance graph representing activities conducted by exactly one true identity, be it the attacker or a normal user.}

The benefit is even more apparent during intra-machine provenance tracing. An enterprise-level application server typically hosts more than a hundred logon sessions at the same time, with some belonging to the same user, and runs cumulatively thousands of logon sessions after just one day operation. For instance, in the Oilrig~\cite{oilrigEP} dataset, system activities run on the Exchange Server spread over more than 5 thousands logon sessions. While some logon sessions are (interactive) long-running ones, many others die soon after a short execution, e.g., accessing some resources. Oblivious to logon session ID, prior provenance tracing approaches would connect all these system activities, as they all at the end can be traced back to the root process, \texttt{system} (PID=4) on Windows. In contrast, our approach identifies the truly involved logon sessions and tracks system activities only inside them, greatly easing the dependency explosion problem.

The efficiency introduced by leveraging logon session ID shows not only during provenance tracing, but also for log storage reduction. Under the assumption of OS integrity, most system activities run under logon sessions with predefined logon ID belong to standard routines, and can be considered as forensics-irrelevant and safely discarded. In the example of the Exchange Server in the Oilrig dataset, these system activities account for over 90\% of all log events created on it. However, some of these system activities are forensics-relevant, as they are related to remote access and logons. \Sys identifies and tracks these system activities during its logon session linking. Further, as logon sessions also introduce the separation of privileges, \Sys automatically identifies privilege escalation during intra-machine cross-session tracing by checking the value of "Token Elevation Type" and "Integrity Level" associated with a logon session ID. By doing so, it has identified, e.g., in Figure~\ref{fig:OilrigAttackGraph}, the privilege escalation from unprivileged user to Administrator, i.e., from \texttt{0xb4874} to \texttt{0xb4829}, and then privilege escalation from Administrator to System, i.e., from \texttt{0xb4829} to \texttt{0xb247f9c3e}.

\subsubsection{Challenges}
We encountered several challenges while developing \Sys. First, each network authentication \& logon process results in multiple logon events with distinct logon session ID on the accessed machine. Depending on the remote access type, e.g., via RDP or SMB (Server Message Block), the number of logon sessions created varies. Second, under certain circumstances, system activities resulted from a new logon are recorded with an existing session ID, leading to false dependency. Third, it is often not possible to reveal the remote access type by inspecting the logon event alone.

\subsubsection{Remote access type inference}
Identifying remote access type is critical for logon session ID reassignment and session linking, both crucial for accurate cross-machine tracing. The way how exactly Windows emits logon events and assigns logon ID is obscure. Due to Windows' closed-source nature, revealing this via source code is not possible. Reference to the most detailed sources about Windows~\cite{Allievi2022,Russinovich2016,Miroshnikov2018,Forshaw2024} also failed to deliver an answer. Hence we resorted to extensive testing and analysis of each remote access type. 

We found that both authentication \& logon events and system activities events need to be processed to extract a list of attributes for accurately inferring remote access type. Hence, for each logon event, \Sys checks related authentication \& logon events and system events to analyze the context and then classify the remote access type. \Sys can identify all standard remote access types: RDP, SSH, Powershell-Remoting (WinRM), WMI, RPC, PsExec, Network Share (SMB), internal web-server request. Note that only identity-based remote access types are of interest, in which valid accounts are used to access a computer. Remote access via malware/C2 agent is not identity-based, and will not create new logon sessions. Repetitive accesses via the same malware/C2 agent run in the same logon session and their system activities are recorded with the same session ID. That is, as long as \Sys can trace to the session in which the C2 agent is present, it can detect all system activities conducted by it, without the need for inferring remote access type and session linking etc. For example, in Figure~\ref{fig:OilrigAttackGraph}, the attacker has performed several attack steps via the C2 agent on the Workstation\_1 at different times, and all of these steps are detected by \Sys, as they are all executed in the same session \texttt{0xb0db1} and \Sys has successfully traced back to this session.

We also found that some logon events and associated logon sessions are purely byproducts. Some of these sessions terminate soon after being created and others live until a logoff occurs. These logon sessions typically do not contain any forensics-relevant system activities. Although their creations are resulted from a user logon, they are not influenced by the logged-on user. For instance, when a domain user logs into a machine via SSH, a byproduct logon session under a virtual user $sshd\_xxx$ gets created.

\subsubsection{Logon session ID reassignment}
For several remote access types, system activities performed on a new logon session are recorded under an existing logon session ID, necessitating session ID reassignment for causally correct tracing. A typical example is RDP, which is often misused by attackers~\cite{T10211}. Unlike SSH, a remote logon session via RDP in the accessed machine will not terminate, when the client program on the accessing machine exits. The user has to explicitly log out to terminate that session. However, users typically close the client program, unaware of their still-running logon sessions on the remote machine. 

When the user's credentials are stolen by an attacker who then logs into the same remote machine with the same credentials, all system activities conducted by the attacker are labeled with the ID of the user's long-running session, although a new logon session with a different ID is created. This is due to Fast User Switching~\cite{FastUserSwitching}, which provides users the same setup after reconnecting to an existing local console session or remote desktop session\footnote{Note that logon sessions and local console / remote desktop sessions are two different concepts. Only a few logon sessions \quotes{live} in local console / remote desktop sessions, which could be seen as a container for one or a few logon sessions.}. The new logon session starts with the new logon by the attacker, and ends when the attacker disconnects (not necessarily explicitly logs off), marking the exact timespan of system activities conducted only by the attacker. Note that it is not possible for the attacker and the user to be on the same remote desktop session (and hence the same logon session) at the same time, a restriction enforced on Windows OS.  Hence, if \Sys identifies a logon as RDP access, it further checks for another specific related event indicating whether the user has entered a new remote desktop session or an existing one. If an existing remote desktop session (and hence the logon session) is being reentered, \Sys takes the ID of the new logon session and replaces the existing logon session's ID (inside the remote desktop session) with it for the corresponding system activities.

Other remote access types requiring session ID reassignment include Powershell-Remoting and internal web-server request. In Figure~\ref{fig:OilrigAttackGraph}, the attacker from the Workstation\_1 leveraged web-shell to execute commands on the Exchange Server. However system activities caused by the attacker's web requests are recorded with System account's logon session ID, i.e., \texttt{0x3e7}, along with system activities resulted from other domain users' web requests. \Sys correctly identified the remote access type and reassigned the attacker's system activities with \texttt{0x2476c741}, separating them from unrelated system activities caused by other users.

\subsubsection{Logon session linking}
Although most system activities under predefined logon ID are forensics-irrelevant, some are directly responsible for the creation of users' logon sessions. For instance, when a user from one machine executes remote commands on another machine via Powershell-Remoting, the remote commands are executed by the process \texttt{wsmprovhost.exe} on the second machine, which is spawned by an instance of \texttt{svchost.exe} process under the System logon session, i.e., \texttt{0x3e7}, once the user successfully authenticated against a domain controller. \Sys links users' logon sessions with those predefined logon sessions, and performs only backward-tracing in those sessions until finding the root process responsible for the creation of the user logon session of each access type. Logon session linking is also needed when a privileged user logs in a remote machine via RDP. Due to UAC~\cite{uac}, two logon sessions are created inside a remote desktop session, of which one has a privileged access token and the other does not. System activities run under both logon sessions can only result from the same true identity. Hence these two sessions need to be linked for accurate tracing.  

\subsection{Threat Score Assignment}

With our summarization of AD attacks in Figure~\ref{fig:ADMapping}, we identified two key insights for accurate AD attack detection. The first insight is that attacks against AD follow a rigid pattern: AD discovery, credential access, lateral movement and privilege escalation. The second insight is that the key enabler of an AD attack is successful credential access, and the most observable result of an AD attack is lateral movement. For instance, in Kerberoastig attacks, a service must be first identified, whose credentials will be then stolen, and used for lateral movement and privilege escalation afterwards.   

Based on these two critical insights, \Sys calculates a threat score for each provenance attack graph generated after logon session-based backward \& forward tracing. This is to ensure that the most likely true attacks are always investigated first by security analysts. For each potential lateral movement, i.e., a cross-machine edge in an attack graph, \Sys checks how many credential access techniques were executed in the corresponding hosts before, and how often each technique was performed. For each performed credential access technique, \Sys explicitly checks whether the corresponding process has accessed the system process \texttt{lsass.exe}, which is the most critical process for managing credentials like password hashes and access tokens on Windows OS, and is hence often abused by attackers. Credential access techniques involving accessing \texttt{lsass.exe}'s memory should be considered more severe. Then, \Sys examines how many AD discovery techniques were operated on related hosts and the frequency of each discovery step conducted. Also, \Sys inspects whether and how many times privilege escalation is observed in the provenance attack graph. 

We formulate the threat score calculation in the following two equations.

\begin{equation}
  \label{eq:LmScore}
  TS_{E} = \sum_{i=1}^{n} (Freq({teq_i}) \times Var({tac_i}))^{w_i} 
\end{equation}

where $n$ denotes the number of tactics involved, e.g., credential access. $Freq({teq_i})$ denotes the accumulated execution frequency of techniques in a given tactic, and $Var({tac_i})$ denotes the number of techniques being applied for the given tactic. Considering both the intensiveness $Freq({teq_i})$ and the extensiveness $Var({tac_i})$ of a tactic being performed aligns with the fact broadly observed in threat reports~\cite{AttCampaigns} that attackers often apply multiple techniques within the same tactic to boost their chances of success. Besides, $w_i$ is a weighting factor introduced for each tactic. Credential access is considered as the most relevant tactic, and receives a higher weight than discovery and privilege escalation, as per our second insight.

Whereas Equation~\ref{eq:LmScore} calculates the threat score for each cross-machine edge, Equation~\ref{eq:SumScore} accumulates the threat scores returned from Equation~\ref{eq:LmScore}, and outputs the threat score for the attack graph.

\begin{equation}
  \label{eq:SumScore}
  TS_{G} = \sum_{i=1}^{n} (TS_i \times (DA + 1) \times Criticality)
\end{equation}

where $n$ denotes the number of cross-machine edges found in a given attack graph, and $TS_i$ denotes the threat score assigned for each cross-machine edge from Equation~\ref{eq:LmScore}. $DA$ indicates whether credentials of domain administrators are involved in the cross-machine system activities, and takes the value 0 or 1. By doing so, we prioritize attack graphs for further investigation, in which credentials of domain administrators are involved due to more severe consequence of these attacks. Similarly, we assign a $Criticality$ to different attack types. Because, for instance, a Golden-Ticket attack may imply a compromise of the entire domain, whereas a Silver-Ticket attack's scope is limited to certain servers in the domain. If credentials of a non-privileged domain user are involved in a Pass-the-Hash attack, the consequence is even less critical.

\section{Implementation}
We implemented a prototype of \Sys in Python, and deployed it on a 64bit Ubuntu 22.04 OS with 256 GB of RAM and a 32-core processor. This machine also hosts Elasticsearch~\cite{elasticsearch}, to which \Sys interfaces via EQL~\cite{eql}, a query language designed for security purposes. Note that, for enhanced efficiency, \Sys functions as an on-demand tracing system, which searches for relevant events to process and outputs attack graphs only after an authentication anomaly is detected in its first stage. Elasticsearch provides scalable and near real-time search for log data investigation.

In AD, authentication logs are recorded only on domain controllers, whereas logon logs are collected in the accessed hosts. We collect authentication logs from the domain controller and logon logs from each domain-joined host.  Authentication logs and logon logs can be linked with a logon GUID. For system activities, we collect Windows Security~\cite{windowssecurity} logs and Sysmon~\cite{sysmon} logs; both are industry-standard logs. To ensure log integrity, recent Sysmon versions start as protected processes, preventing tampering or disabling by attackers. Windows introduced \textit{Protected Event Logging}~\cite{aboutloggingwindows} to secure logs from unauthorized access.

\section{Evaluation}
\label{s:eval}

\PP{Public Datasets} Public datasets often do not contain AD attacks presumably due to higher requirement on setting up emulation infrastructures. For instance, the DARPA E3 dataset~\cite{TransparentComputingE3} does not contain any AD-based attack. Although in the DARPA E5 dataset~\cite{TransparentComputing}, a so-called Copykatz module is used for credential access, the credentials obtained are local credentials, as the corresponding machines are not joined to a domain. The DARPA OpTC dataset~\cite{OpTC} is the only public dataset including AD-based attacks that we can find. Unfortunately, this dataset only includes system logs on domain-joined hosts, but no logs from the domain controller. Authentication logs, which are recorded only in domain controllers, are critical for \Sys's functionality. Hence we cannot evaluate \Sys on the OpTC dataset either. 

\PP{MITRE Attack Emulation} AD attacks are present in nine out of MITRE's eleven full emulation plans~\cite{AdversaryEmulationLibrary}. MITRE Engenuity ATT\&CK\textsuperscript{®} Evaluations~\cite{MITREEngenuity} use these plans to assess commercial detection systems from leading security vendors. MITRE's emulation plans target enterprise networks with more sophisticated, cross-machine attacks than most public datasets, offering greater authenticity. However, MITRE has not published any corresponding datasets. We stringently implemented three emulation plans, i.e, APT29~\cite{apt29EP}, Oilrig~\cite{oilrigEP}, and WizardSpider~\cite{WizardSpiderEP}, and then evaluated \Sys on these datasets. Table~\ref{tab:evaluation:datasetsOverview} gives an overview of our datasets.

\begin{table}[!]
\centering
  \caption{Overview of the evaluation datasets}
  \resizebox{0.48\textwidth}{!}{%
  \begin{tabular}{cccccccc}
    \toprule
    \makecell{Dataset} & \makecell{Target Host\\Number} & \makecell{Ground Truth\\Attack}  & \makecell{Target\\Host OS} & \makecell{Data\\Size}  & \makecell{Event\\Number}  \\
    \midrule
     APT29 & 3 & Golden-Ticket & Windows & 253GB &  160M   \\
    \midrule
     WizardSpider & 3 & Kerberoasting & \makecell{Windows}  & 151GB & 87M  \\
    \midrule
     Oilrig & 4 & Pass-the-Hash & \makecell{Windows}  & 200GB & 116M \\    
    \bottomrule
  \end{tabular}}
  \label{tab:evaluation:datasetsOverview}
\end{table}

\subsection{\Sys vs. SIEM Detection Rules}
We extracted all detection rules for AD-based attacks from the most popular and established SIEM rule repositories, i.e., Elastic~\cite{elasticdetectionrules}, Sigma~\cite{sigma}, Google Chronicle~\cite{chronicledetectionrules}, and converted them into EQL queries, which we then run parallel to \Sys on our datasets. We compare the detection results of Elastic, Chronicle, Sigma and our system \Sys in Table~\ref{tab:evaluation:results}. To ensure fair comparison, in Table~\ref{tab:evaluation:results}, we only show the detection results of these SIEM rules for attack techniques that \Sys's stage 1 can detect (Figure~\ref{fig:AuthenticationIncompleteness}), and not other AD attack techniques like AD discovery, which would lead to a much higher number of false positives.

{\renewcommand{\arraystretch}{1.2}
\begin{table*}[!t]
  \centering
  \scriptsize
  \caption{Comparison of \Sys and prior detection systems.}
  \begin{threeparttable}
  \begin{tabular}{cccccccccccccc}
    \toprule

  \multirow{3}{*}{\textbf{Datasets}}
  & \multicolumn{4}{c }{\Norothead{ \bf \Sys}}
  & \multicolumn{2}{c }{\Norothead{ \bf Elastic}}
  & \multicolumn{2}{c }{\Norothead{ \bf Chronicle}}
  & \multicolumn{2}{c }{\Norothead{ \bf Sigma}}
  & \multicolumn{2}{c }{\Norothead{ \bf CAD}}

  \\ \cmidrule(r{\tbspace}){2-5} 
  
  \multirow{1}{*}{\textbf{}} & \multicolumn{2}{c }{\Norothead{ \bf Stage 1}} & \multicolumn{2}{c }{\Norothead{ \bf Stage 2 }}

  \\ \cmidrule(r{\tbspace}){2-3} \cmidrule(r{\tbspace}){4-5} \cmidrule(r{\tbspace}){6-7} \cmidrule(r{\tbspace}){8-9} \cmidrule(r{\tbspace}){10-11}  \cmidrule(r{\tbspace}){12-13} 
    &  {\bf FP} & {\bf FN}  & {\bf FP} & {\bf FN} & {\bf FP}  & {\bf FN} & {\bf FP} & {\bf FN} & {\bf FP} & {\bf FN} & {\bf FP} & {\bf FN} \\
  
  \midrule
  
  APT29 & \UseMacro{APT29_Scenario2_HADES_Stage1_FP} & \UseMacro{APT29_Scenario2_HADES_Stage1_FN} & \UseMacro{APT29_Scenario2_HADES_Stage2_FP} & \UseMacro{APT29_Scenario2_HADES_Stage2_FN} &  \UseMacro{APT29_Scenario2_Elastic_Lat_Mov_FP} & \UseMacro{APT29_Scenario2_Elastic_Lat_Mov_FN}   &  \UseMacro{APT29_Scenario2_Chronicle_Lat_Mov_FP} & \UseMacro{APT29_Scenario2_Chronicle_Lat_Mov_FN} & \UseMacro{APT29_Scenario2_Sigma_Lat_Mov_FP} & \UseMacro{APT29_Scenario2_Sigma_Lat_Mov_FN} & 0 & 0 \\
  
  \midrule

  WizardSpider & \UseMacro{WizardSpider_HADES_Stage1_FP} & \UseMacro{WizardSpider_HADES_Stage1_FN} & \UseMacro{WizardSpider_HADES_Stage2_FP} & \UseMacro{WizardSpider_HADES_Stage2_FN} & \UseMacro{WizardSpider_Elastic_Lat_Mov_FP}  & \UseMacro{WizardSpider_Elastic_Lat_Mov_FN} &  \UseMacro{WizardSpider_Chronicle_Lat_Mov_FP} & \UseMacro{WizardSpider_Chronicle_Lat_Mov_FN} & \UseMacro{WizardSpider_Sigma_Lat_Mov_FP} & \UseMacro{WizardSpider_Sigma_Lat_Mov_FN} & 0 & 1 \\

  \midrule

  Oilrig & \UseMacro{Oilrig_HADES_Stage1_FP} & \UseMacro{Oilrig_HADES_Stage1_FN} & \UseMacro{Oilrig_HADES_Stage2_FP} & \UseMacro{Oilrig_HADES_Stage2_FN} & \UseMacro{Oilrig_Elastic_Lat_Mov_FP}  & \UseMacro{Oilrig_Elastic_Lat_Mov_FN} &  \UseMacro{Oilrig_Chronicle_Lat_Mov_FP} & \UseMacro{Oilrig_Chronicle_Lat_Mov_FN} & \UseMacro{Oilrig_Sigma_Lat_Mov_FP} & \UseMacro{Oilrig_Sigma_Lat_Mov_FN} & 0 & 1 \\
    
  \bottomrule
  \end{tabular}
    \begin{tablenotes}\footnotesize
     \item[1] Note that each dataset has only one true positive (TP=1).
     \end{tablenotes}
   \end{threeparttable}
\label{tab:evaluation:results}
\end{table*}

}

Note that we implement \Sys as an on-demand PIDS for enhanced efficiency. Only when \Sys's stage 1 detects a potential lateral movement, it runs its stage 2 for a whole-network investigation, which unfolds other tactics like AD discovery and credential access when they are causally related to the lateral movement. That is, \Sys cannot detect an early-stage AD attack, in which the attacker has only performed AD discovery and credential access, but not yet moved to other machines. In particular, for attacks like Golden-Ticket, \Sys's stage 1 cannot detect the creation of a Golden-Ticket, i.e., credential access, but its usage for lateral movement. \Sys's stage 2 checks credential access techniques related to Golden-Ticket, if the stage 1 detects a Golden-Ticket-based lateral movement. \Sys avoids a high rate of false positives by neglecting potential but yet less critical early-stage attack steps.  

Table~\ref{tab:evaluation:results} reveals that, comparing to \Sys, SIEM rules produced more false positives and missed some true attacks. Whereas Chronicle has a relatively low false positive rate in all datasets, it missed the true attack in each dataset. In contrast, Sigma detected all true attacks, at the cost of having a much higher false positive rate. Elastic detected the Pass-the-Hash attack in the Oilrig dataset with less false positives, but missed the attacks in APT29 dataset and WizardSpider dataset. 

To understand why SIEM rules have a high rate of false positives and false negatives, we resort to manually inspect them. On the one hand, we find that a high number of false positives often results from overly general rules matching system activities caused by benign programs often misused by attackers, i.e., LOLBins. Further, Sigma's unusually high number of false positives is also due to the fact that Sigma's rule repository includes many similar rules created by different contributors for the same attack techniques. On the other hand, both Chronicle's and Elastic's rules tend to be overly specific, i.e., having hard-coded strings as conditions, and also often allowlist system activities caused by known \quotes{benign} programs. Simply favoring high-degree of specificity and overly allowlisting lead to less false positives, but inevitably cause more false negatives.  
 
Unlike open-source SIEM rules, \Sys employs an advanced detection strategy that goes beyond solely looking for specific isolated log events, contributing to an average false positive reduction rate of \UseMacro{fpr_avg_reduction}\%. Most activities involved in AD attacks are based on (mis)using legitimate AD infrastructures and services. For example, network shares are used intensively by normal domain users and only occasionally by attackers for lateral movement. That is, attackers can easily blend into the noise caused by legitimate users, as indicative system and network activities related to AD attacks are so prevalent in benign operations. However, when putting truly causally related system activities in a provenance graph via precise cross-machine tracing, it exposes provenance graphs containing system activities conducted by attackers as they typically follow a rigid AD attack pattern unlike benign provenance graphs.   

\begin{figure*}[!t]
  \centering
  \includegraphics[width=0.8\linewidth]{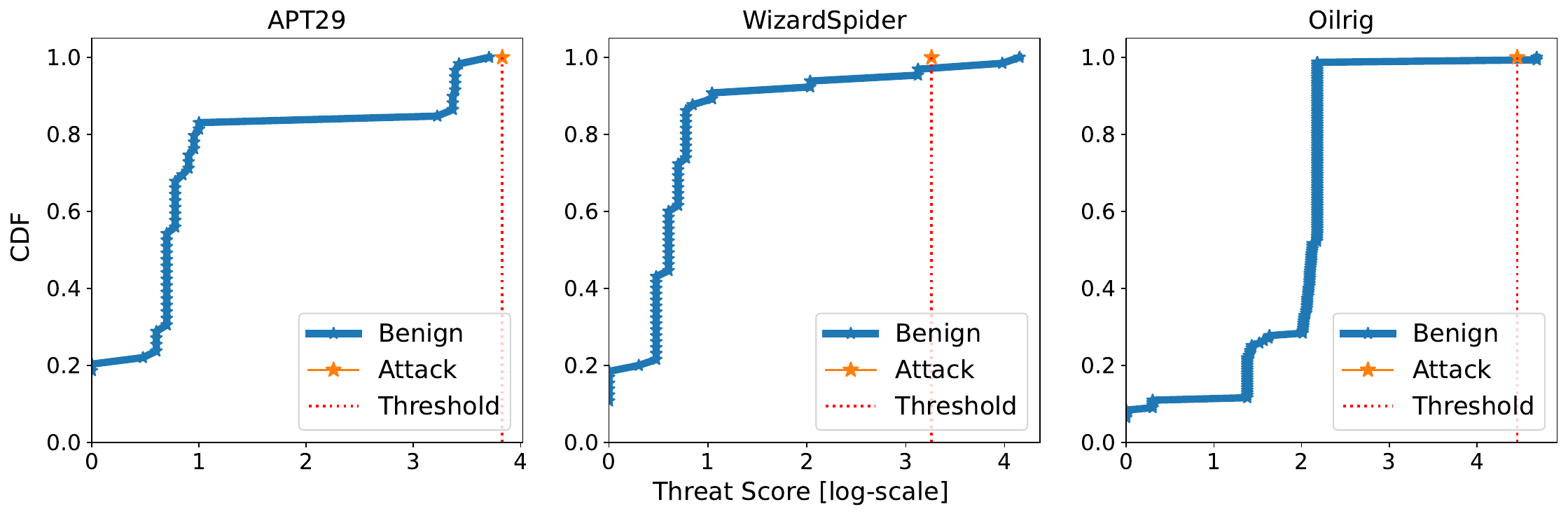}
  \caption{CDF of threat score for false and true alerts.}
  \label{fig:ThreatScore}
  \vspace{-2ex}
\end{figure*}

Detailed comparison between \Sys's stages 1 and 2 is given in Figure~\ref{fig:ThreatScore}, which presents the cumulative distribution function of threat scores for benign and attack alerts. \Sys's stage 2 results in Table~\ref{tab:evaluation:results} are based on the true attack threat score threshold.
Our threat triage algorithm in \Sys's stage 2, combined with accurate cross-machine tracing based on logon session-based execution partitioning, contributes to an average false positive reduction rate of \UseMacro{fpr_avg_1to2_reduction}\% when comparing to its stage 1.

\subsection{\Sys vs. CAD}
Prior research~\cite{Bouwman2020} shows that, contrary to general perception, commercial security products may have poor detection quality despite high price tag. In order to answer the question how much value a commercial attack detection product could bring, we subscribed to a cloud-based dedicated AD attack detection solution from a popular security vendor and evaluated it alongside \Sys on our datasets. The security vendor is considered as a significant security solution provider in the security market overviews of Gartner~\cite{Gartner2023} and Forrester~\cite{Forrester2021}. However we did not get the permission to name the vendor in the present paper, and hence refer to the security product we purchased simply as CAD (commercial attack detector).   

We installed CAD on the domain controller of our emulation infrastructure. We were also provided with a documentation of the security product we purchased, in which a guideline for CAD's logging configuration and its high-level detection logic/rules are included. To our surprise, as shown in Table~\ref{tab:evaluation:results}, the commercial product did not raise a single false alarm, but is plagued with a high false negative rate. The result of our investigation, however, aligns with the findings in prior work~\cite{Bouwman2020}, i.e., commercial security products may sacrifice detection rate in exchange for analyst time.    

By consulting on the documentation, in particular CAD's high-level detection logic,  we were able to find some explanation for the poor detection rate. First, we find that CAD, as a dedicated AD attack detection system, does not collect enough data we consider as necessary for accurate detection.  That is, CAD only collects a limited set of authentication \& logon event types, leading to restricted visibility. Second, some of its detection rules do not fire alerts on events that are previously observed on the same computers. With the prevalence of LOLBins, doing so reduces false positives, but inevitably introduces false negatives.  Third, CAD does not check system logs inside each host at all. We argue that credential access techniques, a critical step in AD attacks, are detectable mostly via system logs.

To avoid false conclusion, we engaged with the vendor over weeks by first checking if our installation and configuration of CAD was correct, and then presenting our finding, and asking for explanation on the high false negative rate. The answer of the vendor confirmed our evaluation findings. In particular, the vendor responded to the false negative for the Pass-the-Hash attack by saying that CAD's detection on Pass-the-Hash attacks has known issues and they are working on a fix.

\begin{figure}[!t]
  \centering
  \includegraphics[width=0.95\linewidth]{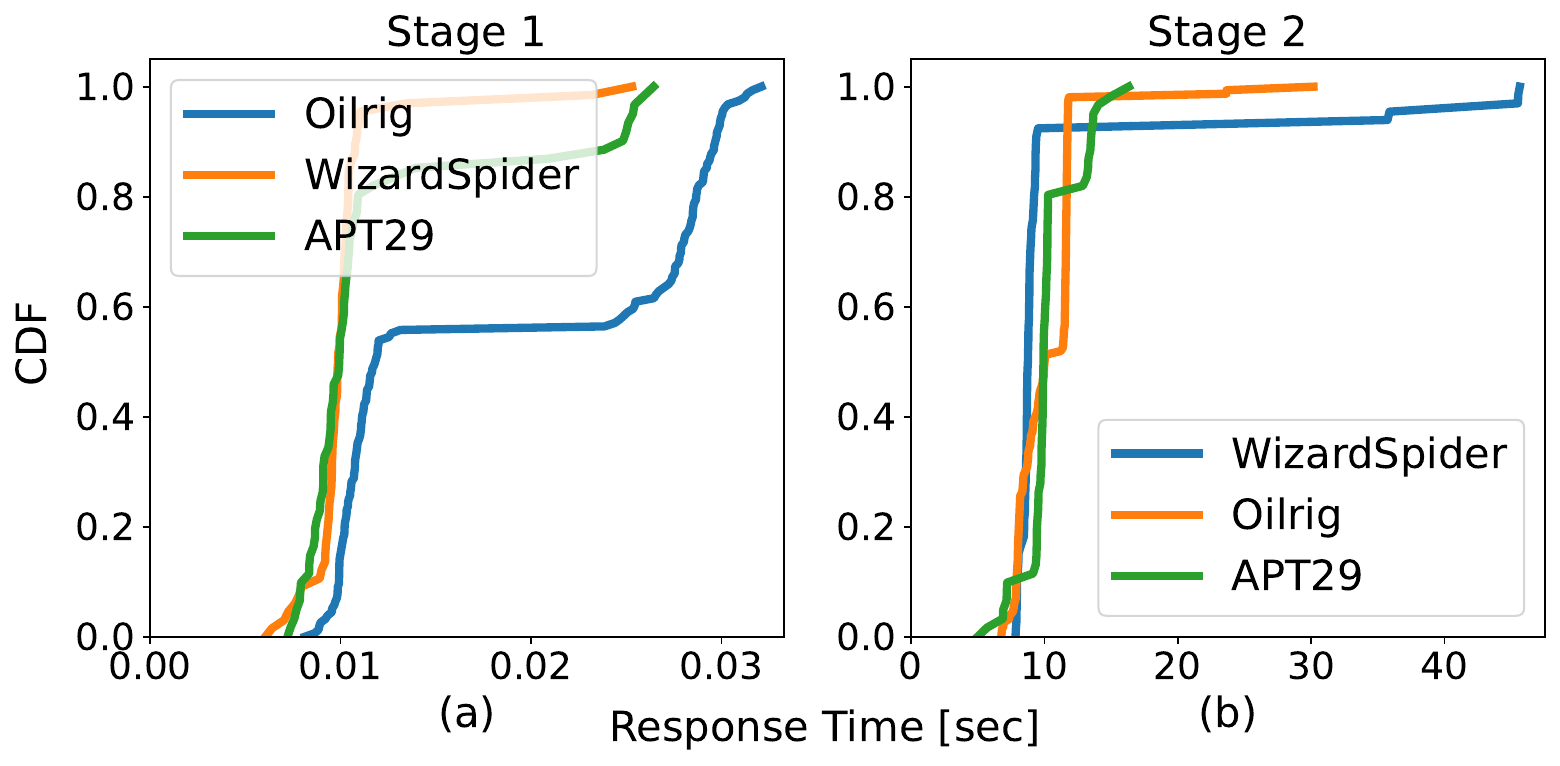}
  \caption{CDF of response time of \Sys.}
  \label{fig:ResponseTime}
  \vspace{-2ex}
\end{figure}

\subsection{Response Time}
\label{ss:ResponseTime}
We measure the response time of \Sys in two separate parts. Figure~\ref{fig:ResponseTime} shows the cumulative distribution function of response time of \Sys in its two stages, respectively. \Sys's stage 1 response time is measured per authentication \& logon alert. Figure~\ref{fig:ResponseTime} (a) reveals that it takes less than 35 milliseconds to find an authentication \& logon anomaly for all authentication \& logon events in each dataset. We measure \Sys's stage 2 response time as the time to perform logon session-based backward \& forward tracing on a high-level alert found in the stage 1, while checking for system activities related to AD discovery techniques, credential access, and privilege escalation, and calculate the threat score for the resulted attack graph.  As shown in Figure~\ref{fig:ResponseTime} (b), it takes at most 45 seconds to return a low-level provenance attack graph with associated threat score for each stage 1's high-level alert in every dataset.

\subsection{Case Study}

\subsubsection{Pass-the-Hash}
Pass-the-Hash attacks leverage NTLM authentication process, instead of the default AD authentication process Kerberos. This authentication anomaly manifests in our authentication anomaly detection model introduced in Section~\ref{s:design}. However, using this model alone introduces many false alerts, as NTLM authentication process is still widely in use in Windows domains. For instance, when a user accesses a remote network share via NTLM, the authentication process against the target server and the domain controller looks the same as in Pass-the-Hash attacks. That is, without resorting to system logs, it is difficult to reliably detect Pass-the-Hash attacks. This is also why CAD missed this attack in the Oilrig dataset. However, we show that, by using whole network provenance tracing and an alert triage algorithm, \Sys can reliably detect Pass-the-Hash, and drastically reduce false positives caused by benign user activities. Figure~\ref{fig:OilrigAttackGraph} automatically created by \Sys presents the entire attack chain conducted by the attacker. 

\subsubsection{Golden-Ticket}
\begin{figure}[!t]
  \centering
  \includegraphics[width=0.99\linewidth]{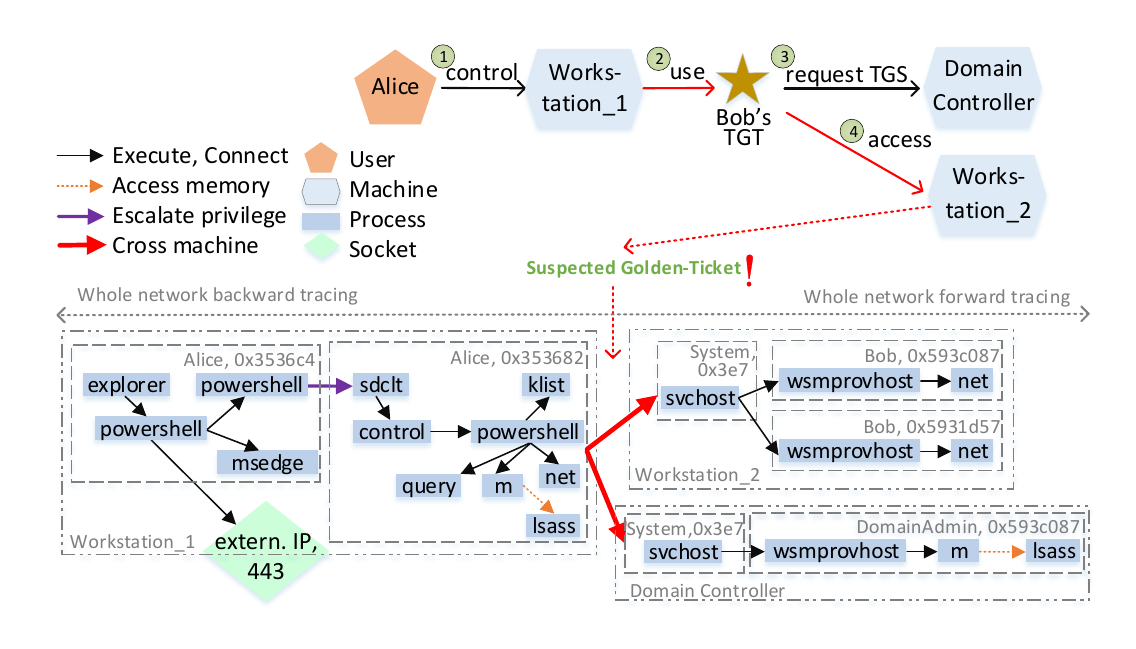}
  \caption{
  An attack graph created by \Sys on the APT29 dataset.}
    \vspace{-3ex}
    \label{fig:Apt29AttackGraph}
\end{figure}

In a Golden-Ticket attack, the attacker manages to steal the credentials of the \texttt{krbtgt} account from a domain controller, and is then able to create a TGT impersonating any domain user. The attacker does not request a TGT from a domain controller before requesting a TGS to an application server, showing an anomaly in our authentication anomaly detection model. However, during a benign operation, when a cached TGT is used to request access to a server, it exhibits the same network authentication anomaly, making Golden-Ticket difficult to detect without inspecting system logs on each involved machines. For the Golden-Ticket attack in the APT29 dataset, as illustrated in Figure~\ref{fig:Apt29AttackGraph}, \Sys first creates an initial high-level attack graph involving AD entities like users and hosts, after it detects an authentication \& logon anomaly and suspects a Golden-Ticket attack. Then it performs system-level forward tracing inside the specific logon session of involved username Bob in the accessed host Workstation\_2, and system-level forward \& backward tracing inside the logon session of Alice in the accessing host Workstation\_1, leading to a logon session in the Domain Controller. This attack graph discloses that the attacker first performed AD discovery and credential access on the initially compromised machine Workstation\_1, then moved to the Domain Controller for further credential access, and finally created a Golden-Ticket for accessing the Workstation\_2.

\section{Discussion \& Related Work}

\subsection{Provenance-based IDS}
Prior PIDS~\cite{hossain2017sleuth, winnower2018,holmes2019,nodoze2019,rapsheet2020,hossain2020combating,KAIROS,FLASH,shadewatcher,PROGRAPHER} are restricted to intra-machine tracing, and unaware of the network context, hindering their adaptability for AD attack detection. That is, by construction, these systems would fail to detect AD attacks. TRACE~\cite{Trace2021} is an early attempt to detect cross-machine attacks in enterprise networks by connecting intra-machine provenance graphs whenever there is a network connection between two machines, oblivious to the domain context. This naive approach inevitably leads to cross-machine dependency explosion, as discussed in Section~\ref{s:design}. In contrast, \Sys is designed with realistic multi-users computing environment in domain context in mind, and can narrow the cross-machine tracing down to a specific logon session containing system activities conducted truly by the same identity, drastically reducing false dependencies. However, \Sys's cross-machine tracing is limited to AD environment and domain-joined machines. Devices brought by employees themselves, aka BYOD, are out of \Sys's scope, since for these devices the authentication \& logon process is done locally with local credentials instead of domain credentials. Yet, local credentials cannot be used in identity-based attacks for moving laterally to other machines in a network.

\subsection{Dependency Explosion}
Intra-machine dependency explosion happens for long-running processes, in which each input event is conservatively considered causally responsible for all subsequent output events, and vice versa, resulting in excessive amounts of false dependencies and formidably dense provenance graphs. Similarly, cross-machine dependency explosion occurs if edges are built simply on a (coarse-granular) network connection basis. Program execution partitioning techniques~\cite{beep,ma2015accurate,protracer} are the first proposed to combat intra-machine dependency explosion. MPI~\cite{mpi2017} improves these techniques by introducing a semantics aware program annotation and instrumentation technique. Whereas all these techniques focus on execution of individual long-running processes, logon session-based execution partitioning operates at a higher level, beneficial for alleviating both intra-machine and cross-machine dependency explosion.  Nonetheless, the existing techniques are complementary to \Sys and can further reduce false dependencies in provenance graphs.

\subsection{Lateral Movement Detection}
Although \Sys is more than a lateral movement detector, its design is centered on lateral movement detection. That is, \Sys's stage 1 alerts on potential lateral movements, which are then further investigated in its stage 2 via whole-network provenance tracing, and triaged with the insights from an extensive analysis of AD attacks conducted by APT actors, accurately and comprehensively unraveling attackers' traversal inside enterprise networks. \Sys overcomes several inherent weaknesses, in terms of logging and assumptions, of state-of-the-art lateral movement detection system Hopper~\cite{ho2021hopper}. First, Hopper proposes an inference algorithm to identify complete login paths in enterprise networks, as leveraging standard authentication logs alone is insufficient to identify those paths. In contrast, \Sys combines authentication \& logon logs and system logs to trace these paths on a logon session basis, producing login paths guaranteed to be correct, unlike Hopper's inference algorithm. Second, Hopper's detection algorithm operates on two assumptions: 1) attackers use a different set of credentials when moving to other machines, 2) attackers eventually access a machine which they previously could not access. While intuitive, these assumptions are not reliable enough, as \Sys's stage 1 shows that detection based on authentication \& logon anomaly alone leads to many false positives. A thorough investigation via detailed system activities checking at \Sys's stage 2 is critical for reducing these false alarms.

\section{Conclusion}
We present a novel accurate AD attack detection system named \Sys in this paper.  Based on a thorough study of AD attacks launched by APT actors, we create a succinct and contextualized AD attack overview revealing key steps and prerequisites of various AD attack types. We leverage critical insights from our extensive analysis on AD attacks to design a light-weight authentication anomaly detection model, and a threat triage algorithm. We propose the concept logon session-based execution partitioning and tracing, which significantly reduces false dependencies, contributing to accurate and swift AD attack detection.  Our evaluations, conducted on datasets derived from rigorously implemented MITRE emulation plans, demonstrate that \Sys significantly outperforms open-source SIEM detection rules and a commercial dedicated AD attack detector.

\balance

\section*{References}
\printbibliography[heading=none]

\end{document}